\documentclass[twocolumn]{aastex62}

\usepackage{bm}

\received{}
\revised{}
\accepted{}

\submitjournal{The Astrophysical Journal}

\shorttitle{Turbulence Dynamo in Galaxy Clusters}
\shortauthors{Roh et al.}

\begin{document}

\title{Turbulence Dynamo in the Stratified Medium of Galaxy Clusters}

\author{Soonyoung Roh}
\affiliation{Department of Physics, School of Natural Sciences, UNIST, Ulsan 44919, Korea}
\author[0000-0002-5455-2957]{Dongsu Ryu}
\affiliation{Department of Physics, School of Natural Sciences, UNIST, Ulsan 44919, Korea}
\author[0000-0002-4674-5687]{Hyesung Kang}
\affiliation{Department of Earth Sciences, Pusan National University, Busan 46241, Korea}
\author{Seungwoo Ha}
\affiliation{Department of Physics, School of Natural Sciences, UNIST, Ulsan 44919, Korea}
\author{Hanbyul Jang}
\affiliation{Department of Physics, School of Natural Sciences, UNIST, Ulsan 44919, Korea}
\correspondingauthor{Dongsu Ryu}
\email{ryu@sirius.unist.ac.kr}

\begin{abstract}

The existence of microgauss magnetic fields in galaxy clusters have been established through observations of synchrotron radiation and Faraday rotation. They are conjectured to be generated via small-scale dynamo by turbulent flow motions in the intracluster medium (ICM). Some of giant radio relics, on the other hand, show the structures of synchrotron polarization vectors, organized over the scales of $\sim$ Mpc, challenging the turbulence origin of cluster magnetic fields. Unlike turbulence in the interstellar medium, turbulence in the ICM is subsonic. And it is driven sporadically in highly stratified backgrounds, when major mergers occur during the hierarchical formation of clusters. To investigate quantitatively the characteristics of turbulence dynamo in such ICM environment, we performed a set of turbulence simulations using a high-order-accurate, magnetohydrodynamic (MHD) code. We find that turbulence dynamo could generate the cluster magnetic fields up to the observed level from the primordial seed fields of $10^{-15}$ G or so within the age of the universe, if the MHD description of the ICM could be extended down to $\sim$ kpc scales. However, highly organized structures of polarization vectors, such as those observed in the Sausage relic, are difficult to be reproduced by the shock compression of turbulence-generated magnetic fields. This implies that the modeling of giant radio relics may require the pre-existing magnetic fields organized over $\sim$ Mpc scales.

\end{abstract}


\section{Introduction} \label{sec:intro}

The baryonic matter in galaxy clusters resides mostly in the intracluster medium (ICM), in the form of hot diffuse plasma. The ICM is known to be highly dynamic and turbulent \citep[][and references therein]{bj2014}. In addition, the ICM contains the magnetic fields of $\mu$G-level, corresponding to the energy density of the order of $\sim1\ \%$ of the thermal energy density, throughout the whole volume of clusters, as revealed in observations of diffuse synchrotron emissions from radio halos and relics and Faraday rotation measures \citep[e.g.,][]{gf2004,fggm2012}.

Magnetic fields play important roles in the ICM, particularly governing microphysical processes, such as the turbulent acceleration of cosmic rays (CRs) and the formation of shock waves, as well as thermal conduction and kinetic viscosity \citep[e.g.,][]{bl2007,kunz10,guo2014,roberg2016,ha2018b}. Yet, the nature and origin of cluster magnetic fields have not yet been fully understood. It was suggested that ``turbulence dynamo'' may be responsible for the generation of ICM magnetic fields, at least in the outskirts \citep[e.g.,][]{ryu2008,porter2015}. Turbulence dynamo, which refers to the amplification of weak seed magnetic fields by turbulent flow motions, generates random fields on scales smaller than the driving scale of turbulence, and hence is often called small-scale dynamo \citep[e.g.,][]{batchelor1950,kazantsev1968,cho2000,schekochihin2004,cho2009}. Typically, it goes through three stages, the initial exponential growth when the magnetic field is dynamically negligible on all scales, the follow-up linear growth when the magnetic energy becomes comparable to the kinetic energy at the dissipation scale, and finally the saturation stage when the magnetic energy accounts for a substantial fraction of the turbulent energy.

Seed fields for cluster magnetic fields, however, are unknown and varied, although many candidates, ranging from primordial to plasma physical, and astrophysical, have been suggested \citep[e.g.,][for reviews]{ryu2012,widrow2012}. Fermi-LAT observations of blazars set a lower bound of $\sim10^{-16}-10^{-15}$ G at the scale of $\sim1$ Mpc for void magnetic fields \citep[e.g.,][]{neronov2010,tavecchio2010}. Planck observations of cosmic microwave background (CMB) anisotropies put a strong upper limit of $B\lesssim10^{-9}$ G again at the scale of $\sim1$ Mpc (comoving strength and scale) for the primordial field strength \citep{planck16}. Hence, the initial seed fields for cluster magnetic fields would be many orders of magnitude weaker than the observed fields of $\sim\mu$G strength.

Turbulence, and hence ensuing dynamo, in galaxy clusters differ significantly from those in other astrophysical environments, such as molecular clouds and star-forming regions. For instance, the ICM turbulence is induced in highly stratified backgrounds, and driven sporadically by mergers during the hierarchical formation of the large-scale structure (LSS) of the universe \citep[e.g.,][]{miniati15,vazza2017}. In addition, while supersonic turbulence is common in astrophysical environments, the ICM turbulence is expected to be subsonic with turbulence Mach number $M_{\rm turb}\sim1/2$ or so \citep[e.g.,][]{ryu2008,porter2015}.

Magnetic field amplification in the ICM have been studied by several authors, using cosmological magnetohydrodynamic (MHD) simulations for the LSS formation \citep[e.g.,][]{vazza2014,marinacci2015,vazza2018}. In turbulence dynamo, the initial exponential growth is rapid, so the memory of seed field properties, such as the strength and length scale, is supposed to be lost. However, the aforementioned studies using cosmological MHD simulations so far have shown that the amplification of magnetic fields due to turbulence dynamo alone (without radiative cooling and feedback processes) is limited. Hence, in order to amplify seed fields via turbulence dynamo to the observed strength of $\sim\mu$G in clusters, the void fields have to be around nanogauss, which is stronger than that inferred from other observations mentioned above. This indicates that either turbulence dynamo may not be the main mechanism that amplifies the ICM magnetic fields, or these cosmological simulations may not have fully reproduced turbulence dynamo in the ICM, perhaps owing to insufficient grid resolution.

In the outskirts of galaxy clusters, the so-called radio relics, elongated structures of diffuse synchrotron emission, have been observed \citep[e.g.,][for reviews]{bruggen12,fggm2012}. They are thought to be associated with ICM shocks of Mach number $M_s\sim 2-3$, induced by mergers \citep[e.g.,][for review]{vanweeren2019}. The synchrotron radiation is polarized with average polarization fraction $\sim10-30\%$, and the polarization vectors are aligned with the shock normal. A spectacular example is the Sausage relic in cluster CIZA J2242.8+5301 \citep{vanweeren2010}. It shows a very high polarization faction of $\sim50-60\%$ with highly aligned polarization vectors over its length of $\sim2$ Mpc. These radio relics may indicate the presence of the magnetic fields coherent over $\sim$ Mpc scales in the cluster outskirts, possibly challenging the small-scale, turbulence dynamo origin of cluster magnetic fields.

In this paper, we investigate whether turbulence dynamo could be the mechanism for the generation of observed magnetic fields in galaxy clusters. We first check whether the cluster magnetic fields of $\sim\mu$G strength could emerge from very weak seed fields, consistent with the inferred lower bound of the void fields. We also examine whether the structures of synchrotron polarization vectors observed in radio relics could be explained by the compression of turbulence-generated magnetic fields across weak ICM shocks.

Toward that end, instead of cosmological structure-formation simulations, we employ turbulence simulations in a controlled periodic volume. To model the realistic cluster environment, the background medium is stratified radially in hydrostatic equilibrium with an external gravity, and the turbulence is driven sporadically mimicking major mergers as described above. Some aspects of turbulence and dynamo in the stratified background were previous investigated. For instance, \citet{shi2018} discussed the behavior of the hydrodynamic turbulence influenced by the buoyancy of the stratified ICM, using cosmological structure-formation simulations. \citet{jabbari2014} examined the magnetic field evolution in the mean-field dynamo of stellar convection. However, the effects of the background stratification on small-scale dynamo by turbulent flow motions have not yet been fully explored and quantified.

\begin{figure*}[t]
\vskip 0 cm
\hskip 0 cm
\centerline{\includegraphics[width=1.05\textwidth]{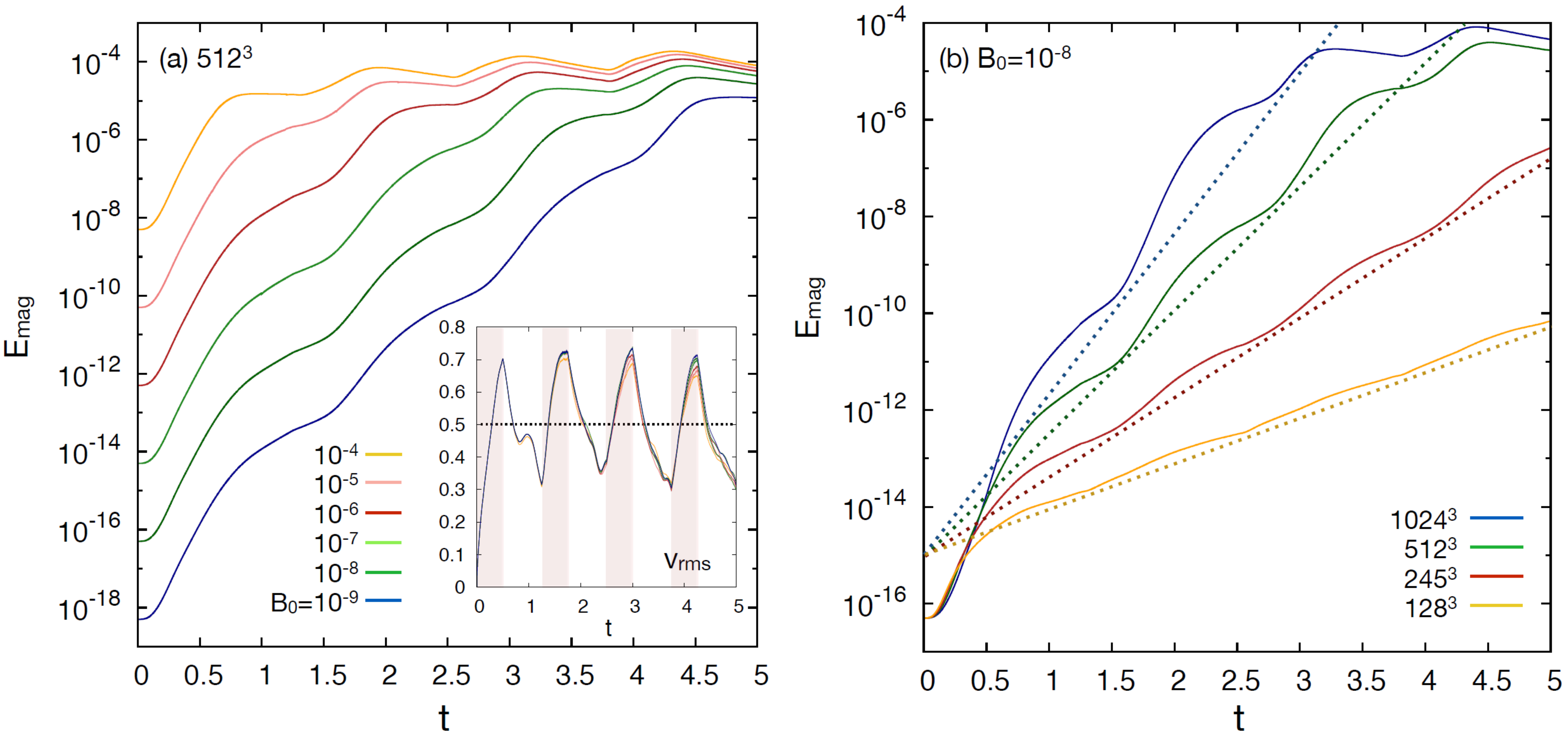}}
\vskip -0.2 cm
\caption{\label{fig:f1}
Magnetic field amplification in turbulence dynamo. (a) The evolution of the magnetic energy, $E_{\rm mag}=(1/2)B^2$, with different seed fields of $B_0=10^{-9}$ to $10^{-4}$ in simulations using $512^3$ grid zones. The evolution of the rms flow velocity, $v_{\rm rms}\equiv\left<v^2\right>^{1/2}$, is shown in the inserted box; the lines of different colors almost overlap. (b) The evolution of $E_{\rm mag}$ with $B_0=10^{-8}$ in simulations using different numbers of grid zones. Dashed lines shows the fittings of the late exponential growth (see the text). Here, $B=1$ is equivalent to $\sim5\times10^{-5}$ G and $t_{\rm end}=5$ to 13 Gyrs, if the model parameters are scaled to the physical parameters relevant for the Coma cluster.}
\end{figure*}

We here employ an MHD simulation code with high-order accuracy, which performs better for the amplification of magnetic field. Our simulation results can be directly compared to previous turbulence simulations, in which the background medium is uniform and the turbulence forcing is continuous in time. And we will test numerically-converged results with sufficient grid resolutions. The paper is organized as follows. Numerical setups and simulations are outlined in Section \ref{sec:simulations}. Results are described in Section \ref{sec:results}. A brief summary follows in Section \ref{sec:summary}.

\section{Simulations}\label{sec:simulations}

We carried out simulations by solving equations for isothermal, compressible MHD flows, where the gas pressure is modeled as $P_g \equiv \rho c_s^2$ with a constant sound speed $c_s$. A three-dimensional (3D) code based on the MHD version \citep{jiang1999} of a weighted essentially non-oscillatory (WENO) scheme \citep{jiang1996} was employed. The eigenvalues and eigenvectors of isothermal MHD flows in \citet{kim1999} were implemented. With a 4th-order Runge–Kutta (RK) scheme for the time evolution, the code has a 5th-order spatial accuracy and a 4th-order temporal accuracy. The $\bm{\nabla}\cdot\bm{B}=0$ constraint was enforced using a constrained transport (CT) scheme \citep{ryu1998}. Viscous and resistive dissipations are not explicitly modeled.

Conventional MHD turbulence simulations adopt a periodic volume in an initially uniform background. Here, we consider a periodic simulation box that contains a spherically symmetric halo with the stratified gas distribution. To emulate the ICM distribution of galaxy clusters, the background medium is modeled by an isothermal $\beta$ distribution \citep[e.g.,][]{cavaliere1976} as follows:
\begin{equation}
\rho(r) = \rho_0\left[1+\left(\frac{r}{r_c}\right)^2\right]^{-3\beta/2},
\end{equation}
where the center of the simulation box is located at $r=0$. The dimensionless gas density and pressure are initialized with $\rho_0=1$ and $P_{g,0}=1$ (so $c_s=1$) in simulation units, respectively. We set $\beta=1$ and $r_c=0.075$. Simulations were performed in a cubic box of dimensionless size $L_0=1$, using $N_g^3=128^3$ to $1024^3$ grid zones. 

\begin{figure}[t]
\vskip 0 cm
\hskip -0.2 cm
\centerline{\includegraphics[width=0.5\textwidth]{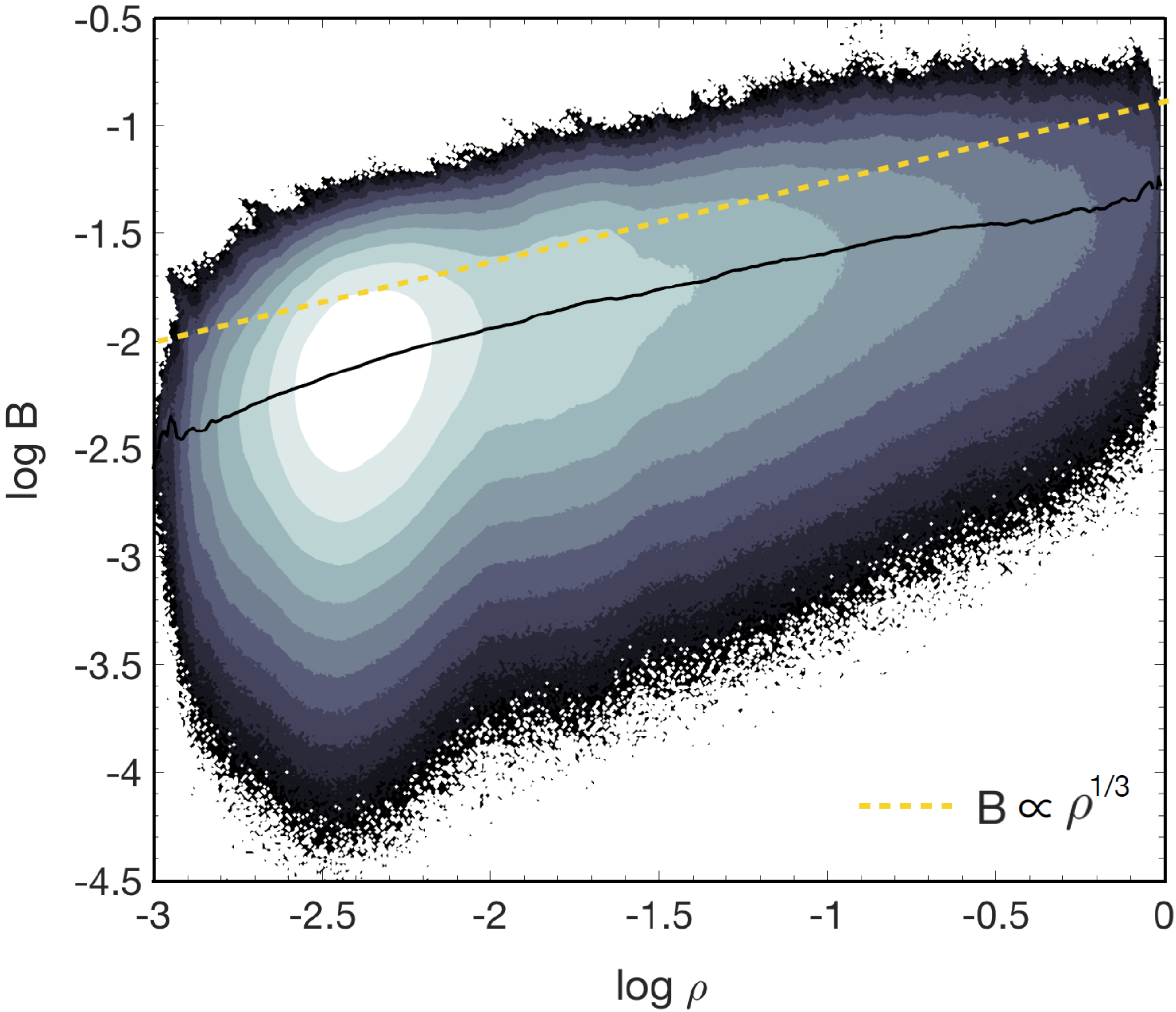}}
\vskip -0.2 cm
\caption{\label{fig:f2}
Correlation between the gas density and the magnetic field strength at $t=4.6$ in the high-resolution simulation of $1024^3$ grid zones with the initial background magnetic field strength of $B_0=10^{-8}$. The black solid line shows the average magnetic field strength, and the yellow dashed line draws the $B\propto\rho^{1/3}$ scaling relation.}
\end{figure}

An external gravity balancing the pressure gradient due to the stratification, $g=c_s^2d\ln\rho/dr$, is added to the momentum equation. Initially, the medium is at rest with $\bm{v} = 0$, and the computational volume is permeated with a uniform magnetic field of $\bm{B}_0=(B_0,0,0)$ parallel to the $\hat{x}-$ axis; a range of $B_0=10^{-9}-10^{-4}$ is considered.\footnote{The unit of $\bm{B}$ is chosen so that the magnetic energy is given as $E_{\rm mag}=(1/2)B^2$.}

Turbulence is driven with ``solenoidal'' forcing, i.e., $\bm{\nabla}\cdot\delta\bm{v}=0$. The velocity perturbations, $\delta\bm{v}$, are drawn from a Gaussian random field of power spectrum, $|\delta v_k|^2 \propto k^6\exp(-8k/k_{\rm inj})$ with $k_{\rm inj}=8k_0$ $(k_0=2\pi/L_0)$, and added to $\bm{v}$ \citep[e.g.,][]{stone1998,maclow1999}. They have random phases, so the driving is temporally uncorrelated. Unlike other conventional turbulence simulations where the forcing is applied continuously in time, the velocity perturbations are enforced ``sporadically'', mimicking major merger events in galaxy clusters. Considering each cluster suffers a few to several major mergers \citep[e.g.,][]{miniati15,vazza2017}, the forcing is turned on for the duration of $\Delta t_{\rm on} =0.5$ and off for $\Delta t_{\rm off}=0.75$, 4 times until the end of simulations at $t_{\rm end}=5$.

The amplitude of the forcing is tuned, so that $v_{\rm rms}\equiv\left<v^2\right>^{1/2}$ is in the range of $\sim0.3-0.7$ with a mean of $\sim1/2$. With $c_s=1$, the mean turbulence Mach number is $M_{\rm turb}\equiv v_{\rm rms}/c_s\sim 1/2$, which is close to the values expected for the subsonic turbulence in the ICM \citep[e.g.,][]{ryu2008,porter2015}.

The inserted box in Figure \ref{fig:f1}(a) shows the development and variation of $v_{\rm rms}$ with our sporadic forcing. Colored and white parts mark the periods during which the forcing is turned on and off, respectively. As shown with different colored lines, the evolution of $v_{\rm rms}$ is insensitive to the strength of the initial background magnetic field, $B_0$, as well as to numerical resolution (not shown).

In the discussion of results below, we consider the Coma cluster as the representative cluster. Considering the core radius of the Coma cluster is $r_c\simeq300$ kpc \citep[e.g.,][]{briel1992,orgrean2013}, the simulation box size is regarded to be $L_0 = 4$ Mpc; then, the grid resolution is $\Delta x\equiv L_0/N_g=3.9-31.25$ kpc. The peak scale of the velocity forcing is close to $\sim L_0/8=500$ kpc, typical size of subclumps in major mergers \citep[e.g.,][]{ha2018a}. We assign $c_s=1.5\times 10^3 ~{\rm km s^{-1}}$, appropriate for the Coma's ICM temperature of $kT=8-9$ keV \citep[e.g.,][]{sato2011}; then the unit time corresponds to $t_0=L_0/c_s=2.6$~Gyrs, and hence our simulations ran up to $t_{\rm end}=13$ Gyrs, close to the age of the universe. Finally, assuming that $\rho_0$ represents the ion number density of $3.5\times10^{-3}$ cm$^{-3}$ at the Coma cluster center \citep[e.g.,][]{briel1992}, the magnetic field unit corresponds to $\sim5\times10^{-5}$ G, and hence $B_0$ is equivalent to $\sim5\times10^{-14}-5\times10^{-9}$ G.

\section{Results}\label{sec:results}

\subsection{Amplification of Seed Magnetic Fields}\label{sec:3.1}

Figure \ref{fig:f1} illustrates turbulence dynamo in simulations with different initial $B_0$ and different numerical resolutions. The left panel shows that the saturation level of the magnetic energy, $E_{\rm mag}$, at $t_{\rm end}$ depends on $B_0$. The right panel demonstrates that the growth of $E_{\rm mag}$ depends rather strongly on numerical resolution. The evolution of $E_{\rm mag}$ looks different from that in simulations for turbulence dynamo in uniform media with continuous driving of turbulence. \citep[see, e.g., Figure 1 of][]{porter2015}, owing partly to the sporadic nature of the forcing and partly to the stratification of the background medium. After the quick initial exponential growth during $t\lesssim0.5$, the follow-up growth is also approximated to be exponential rather than linear. 
This second exponential growth is slower than the initial growth, and hence the final $E_{\rm mag}$ depends on the initial $B_0$ in simulations of $512^3$ grid zones as shown in Figure \ref{fig:f1}(a), and also on the grid resolution as shown in Figure \ref{fig:f1}(b).

If the amplification of magnetic field, $B$, during the second exponential growth stage is modeled as $\left<B\right>\propto\exp(t/\tau_{\rm growth})$ [dashed lines in Figure \ref{fig:f1} (b)], then the growth time scale can be fitted to $\tau_{\rm growth}\approx10^2\Delta x+0.13$ in the computational units. In the physical units appropriate for the Coma cluster, this corresponds to
\begin{equation}
\tau_{\rm growth}\approx0.066\left(\frac{\Delta x}{\rm kpc}\right)+0.34\ {\rm Gyrs}.
\end{equation}
In the highest resolution simulation with $\Delta x=3.9$ kpc ($1024^3$ grid zones), $\tau_{\rm growth}\approx0.6$ Gyrs; hence, $\left<B\right>$ can grow by a factor of $10^7$ during $\sim10$ Gyrs. With the initial seed fields of $B_0\approx10^{-15}$ G, which is close to the lower bound of the void magnetic field (see the Introduction), the mean magnetic field of $\left<B\right>\sim\mu$G would develop within $\sim10$ Gyrs, only if the grid size is sufficiently small, such as $\Delta x\lesssim2$ kpc.

\begin{figure}[t]
\vskip 0 cm
\hskip -0.1 cm
\centerline{\includegraphics[width=0.5\textwidth]{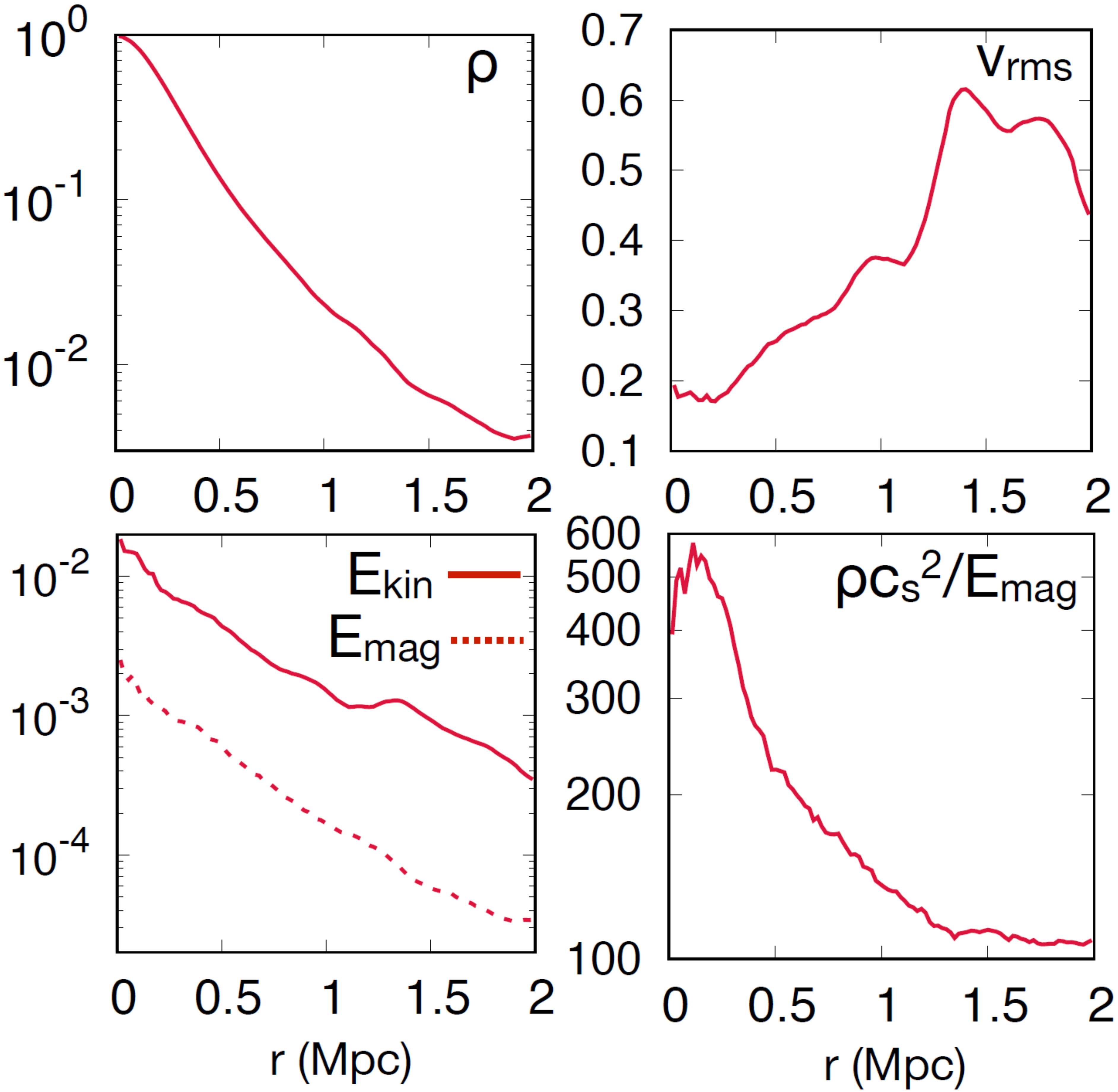}}
\vskip -0.2 cm
\caption{\label{fig:f3}
Radial profiles of $\rho$, $v_{\rm rms}$, $E_{\rm kin}$ and $E_{\rm mag}$, and $(\rho c_s^2)/E_{\rm mag}$ at $t=4.6$ in the high-resolution simulation of $1024^3$ grid zones with the initial background magnetic field strength of $B_0=10^{-8}$. While the quantities are plotted in the simulation units, the radius from the center is given in units of Mpc, adopting $L_0 = 4$ Mpc.}
\end{figure}

The Coulomb mean-free path between collisions is given as $\lambda_{\rm Coul}\sim0.3\ T_{\rm keV}^2/n_{-3}$ kpc, where $T_{\rm keV}$ is the ICM temperature in keV and $n_{-3}$ is the particle number density in units of $10^{-3}$ cm$^{-3}$ \citep[see, e.g.,][]{narayan2001}. It is of order of kpc in the cluster core, while it is as large as tens of kpc in the outskirt. However, it is likely that the effective mean-free path is reduced by plasma instabilities \citep[e.g.,][]{bj2014}. In fact, analyzing the fluctuations in the Chandra data of the Coma cluster, \citet{zhuravleva2019} recently argued that the effective mean-free path would be smaller by two orders of magnitude than the Coulomb value. Our results indicate that if the fluid description of the ICM is valid down to the scale of the order of kpc, turbulence dynamo should be able to amplify the primordial magnetic field of as weak as $\sim10^{-15}$ G to the cluster magnetic field of $\sim\mu$G within the age of the universe.

As described in the Introduction, there have been cosmological simulation studies in which the amplification of magnetic fields via turbulence dynamo was examined in the context of the LSS formation \citep[e.g.,][]{vazza2014,marinacci2015,vazza2018}. So far those simulations failed to reproduce a sufficient amplification of $B$ on cluster scales. This should be partly because their spatial resolution of several kpc is not fine enough, and partly because their numerical codes are only second-order accurate. Moreover, those simulations employed either AMR (adaptive mesh refinement) or moving mesh techniques in order to achieve high resolutions inside clusters. We expect that numerical details, such as the code accuracy and the grid structure, would affect the amplification of $B$. In particular, although not shown here, we found that the high-order WENO code effectively has the resolution enhancement of a factor of two, compared to second-order accurate codes such as the TVD code \citep{ryu1995,kim1999}, when the same number of grid zones is used.

\begin{figure*}[t]
\vskip 0 cm
\hskip -0.1 cm
\centerline{\includegraphics[width=1\textwidth]{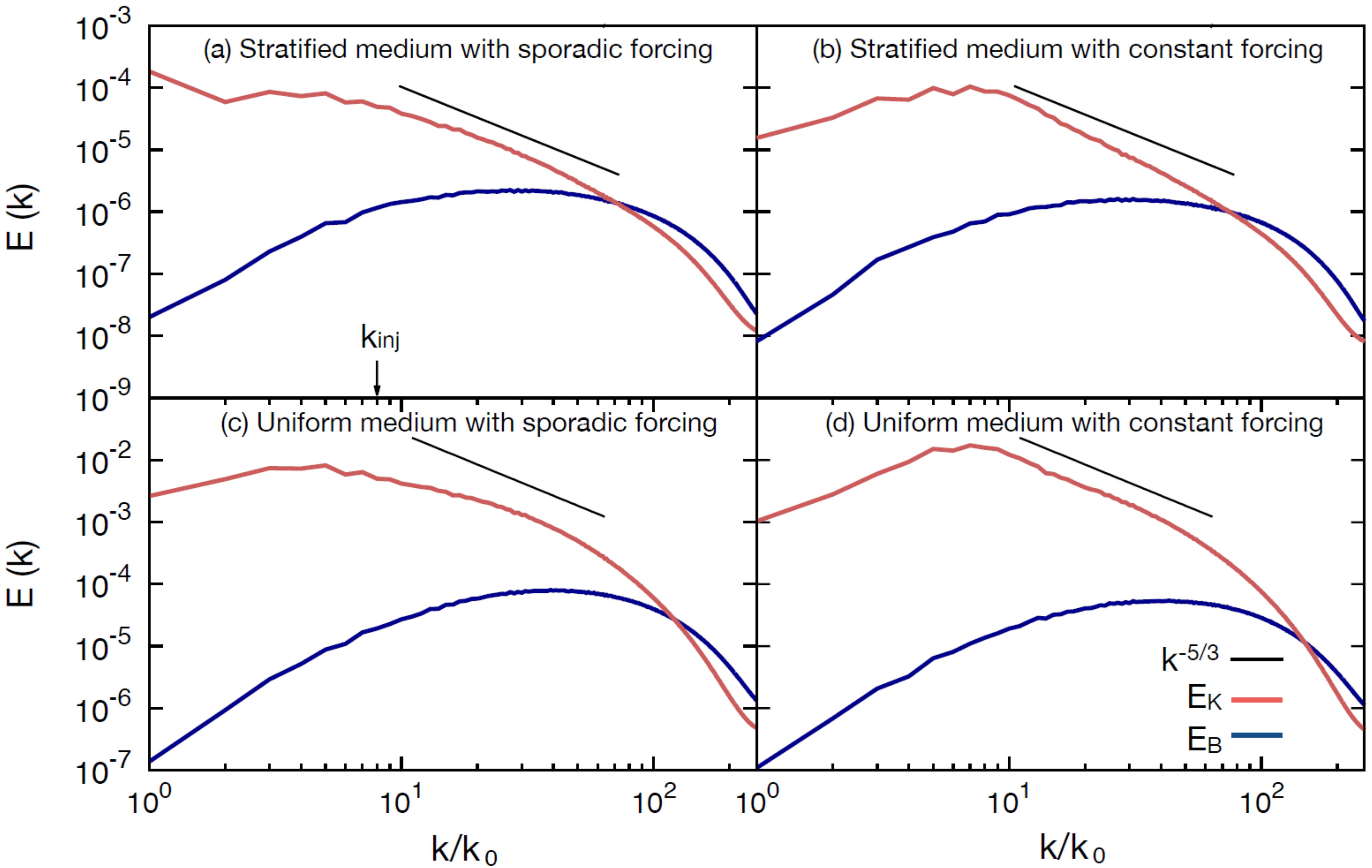}}
\vskip -0.2 cm
\caption{\label{fig:f4}
Comparison of the energy power spectra in turbulence simulations with different setups: (a) in a radially stratified medium with a sporadic forcing, (b) in a radially stratified medium with a continuous forcing, (c) in a uniform medium of initially $\rho=1$ with a sporadic forcing, and (d) in a uniform medium with a continuous forcing. The red and blue lines show the kinetic energy spectrum, $E_K(k)$, and the magnetic energy spectrum, $E_B(k)$, respectively. The black lines draw the Kolmogorov slope. Here, $k_0=2\pi/L_0$. In all the cases, results from the simulations of $512^3$ grid zones with $B_0=10^{-6}$ at $t=4.6$ are presented; $k_{\rm inj}=8k_0$ and $v_{\rm rms}$ of each model are identical.}
\end{figure*}

However, we should note that the current simulations, which were performed in an idealized setup with one model cluster in a periodic box, does not include effects involved in full cosmological simulations, such as the expansion of the universe and realistic mergers during the hierarchical formation of the LSS of the universe. Hence, our conclusion on the capability of turbulence dynamo for the amplification of $B$ in galaxy clusters should be verified in the future with full cosmological simulations of resolution $\Delta x\sim1$ kpc or so, possibly in a (non-AMR) uniform grid structure, using codes of high accuracy. 

Figure \ref{fig:f2} shows the $B-\rho$ relation in the $1024^3$ simulation at $t=4.6$, the epoch when $M_{\rm turb}$ is close to 1/2 [Figure \ref{fig:f1}(b)]. In previous simulations in uniform media with continuous driving, supersonic turbulence, which is relevant for the environment of molecular clouds, gives the scaling relation of $B\propto$ $\rho^\kappa$ with $\kappa=0.3\sim0.5$ \citep[e.g.,][]{padoan1999,ostriker2001}. For subsonic turbulence of $M_{\rm turb}\lesssim1$ with weak seed fields, the correlation between $B$ and $\rho$ is rather weak with small correlation coefficient, although the coefficient value depends on $M_{\rm turb}$ and $B_0$, as well as the details of forcing \citep[e.g.,][]{wu2009,yoon2016}. Cosmological structure-formation simulations, in particular, those without radiative cooling and feedback processes, on the other hand, produced the relation close to $B\propto\rho^{2/3}$ for galaxy clusters \citep[e.g.,][]{vazza2014,marinacci2015}, implying that possibly compression would have played a significant role on the amplification of $B$. Our simulations give a relation close to $B\propto\rho^{1/3}$. We interpret that the $B\propto\rho^{1/3}$ scaling is a consequence of turbulence dynamo, but also affected by the stratification of the background medium.

Figure \ref{fig:f3} shows the radial profiles of $\left<\rho\right>$, $\left<v^2\right>^{1/2}$, $\left<(1/2)\rho v^2\right>$ and $\left<(1/2)B^2\right>$, and $\left<\rho c_s^2\right>/\left<(1/2)B^2\right>$, again in the $1024^3$ simulation at $t=4.6$. The average was calculated over a thin shell at the given radius.  
Although the rms flow velocity increases from the core to the outskirt, the kinetic and magnetic energies decrease due to the density stratification, as also shown in cosmological structure-formation simulations \citep[e.g.,][]{vazza2017,dominguez2019}. 
The decrease of $v_{\rm rms}$ at large radii ($r> 1.3$~Mpc) should be the consequence of the periodic boundary adopted in the simulation.
In the simulation shown here, the magnetic field has amplified to the level of $E_{\rm mag}\sim(1/10)E_{\rm kin}$ throughout the cluster. In the outskirt, the ratio of the effective thermal energy, $E_{\rm th}=\rho c_s^2$, and the magnetic energy (i.e., the plasma beta) approaches $\sim100$ , close to the value expected in the ICM \citep[e.g.,][]{ryu2008,bj2014}.

\subsection{Scale of Magnetic Fields}\label{sec:3.2}

\begin{figure*}[t]
\vskip 0 cm
\hskip -0.1 cm
\centerline{\includegraphics[width=1\textwidth]{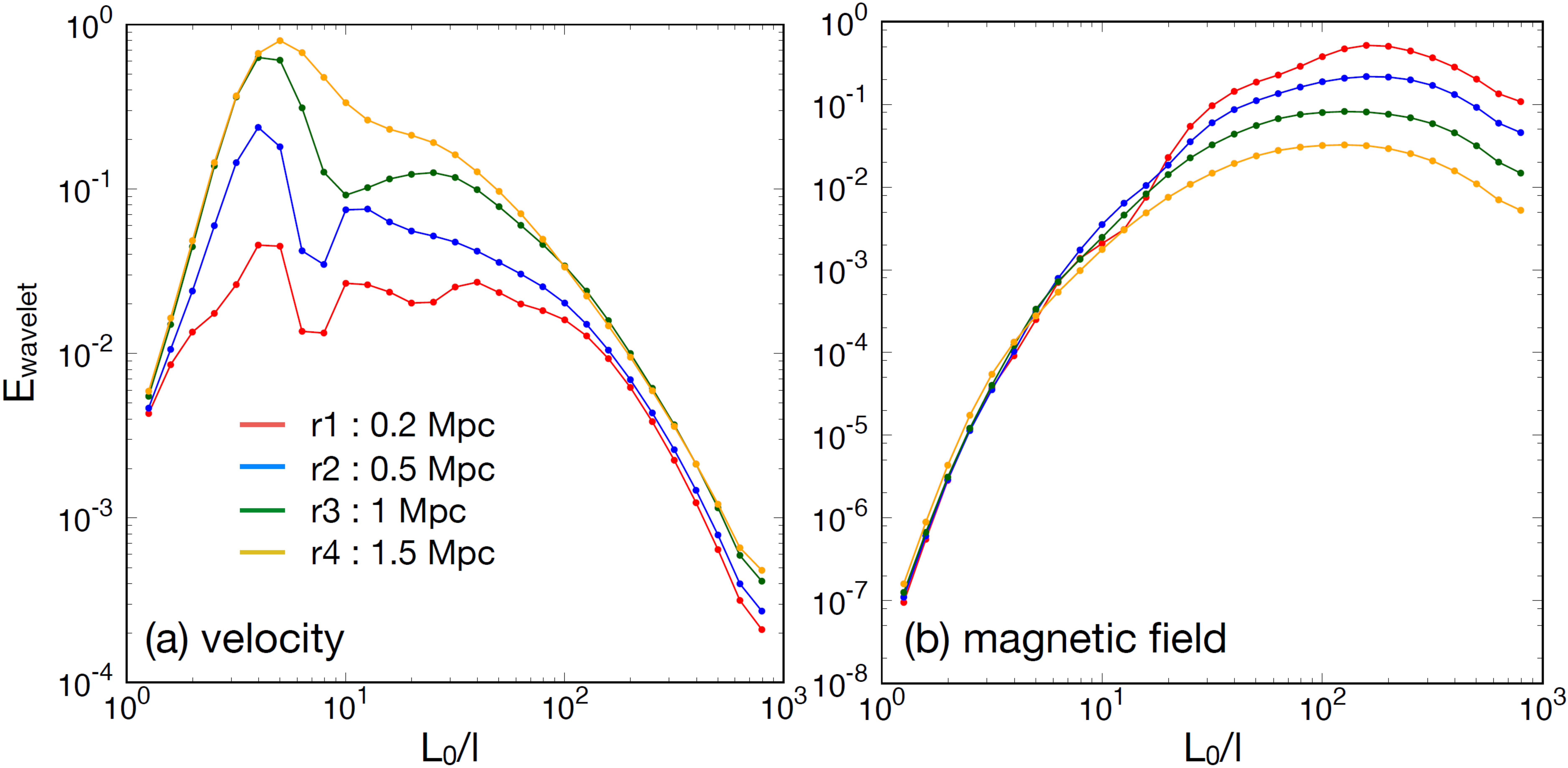}}
\vskip -0.2 cm
\caption{\label{fig:f5}
Wavelet power spectra of the flow velocity (a) and the magnetic field (b) as a function of the effective wavenumber of the filtering scale, $L_0/l$, at four different radii ($r_i=0.2$, 0.5, 1.0, and 1.5 Mpc, shown with different colors) from the center, for the simulation same as in Figure \ref{fig:f4}(a). $L_0 = 4$ Mpc is adopted. The vertical scales of the plots are arbitrary.}
\end{figure*}

The amplification of the strength of $B$ is one aspect of turbulence dynamo, while the growth of the coherence scale of $B$ is another aspect \citep[e.g.,][]{choryu2009}. To examine the scale issue, the kinetic and magnetic energy power spectra from a simulation with $512^3$ grid zones are shown in Figure \ref{fig:f4}(a). They are compared with those for turbulences differently generated: (b) a radially stratified medium with a continuous forcing, (c) a uniform medium with a sporadic forcing, and (d) a uniform medium with a continuous forcing. The injection scale of the forcing, $k_{\rm inj}$, is the same and $v_{\rm rms}$ is tuned to be similar at the time shown, $t=4.6$, in all four models.

In the case of a uniform background with a continuous forcing [Figure \ref{fig:f4}(d)], turbulence exhibits the expected behaviors; that is, while the kinetic energy power spectrum, $E_K(k)$, has the peak around $k_{\rm inj}$, the magnetic energy power spectrum, $E_B(k)$, has the peak at a scale a few times smaller than the injection scale \citep[e.g.,][]{cho2009,porter2015}. With our model for the sporadic forcing, $t=4.6$ corresponds to an epoch of decaying turbulence, and hence the peak of $E_B(k)$ is expected to migrate to a larger scale \citep[e.g.,][]{campanelli2007}. In addition, the stratification of the background medium induces powers in $E_K(k)$ on the cluster scale \citep[e.g.,][]{vazza2014}, and hence the peak of $E_B(k)$ shifts to larger scales, compared to the case with the uniform background. Combining the two effects, the peak of $E_B(k)$ in panel (a) indeed locates at a scale larger than that in panel (d). However, we find that the shift is not dramatic, just a factor of two or so; the peak scale of $E_B(k)$ is at $k/k_0\simeq20$ in panel (a), while at $k/k_0\simeq40$ in panel (d).

Figure \ref{fig:f4}(a) shows that $E_K(k)$ and $E_B(k)$ cross at the scale of $k/k_0\approx 70$ ($60$ kpc for $L_0=4$~Mpc). This crossing has been also seen in some of high-resolution cosmological simulations \citep[e.g.,][]{dominguez2019}. On this scale, the turbulence eddy speed roughly equals the Alfv\'en speed, and hence it is often called the Alfv\'en scale, $l_A$. Below this scale, the Maxwell stress is larger than the Reynolds stress and the flow becomes MHD. The Alfv\'en scale increases slightly with time in our simulations; that is, $l_A$ becomes somewhat larger in later stage. In our model cluster, $l_A$ is larger by an order to magnitude or more than the Coulomb collision scale, $\lambda_{\rm Coul}$, while previously $l_A\lesssim\lambda_{\rm Coul}$ was often expected \citep[e.g.,][]{bj2014}.

\begin{figure*}[t]
\vskip 0 cm
\hskip 0 cm
\centerline{\includegraphics[width=1.05\textwidth]{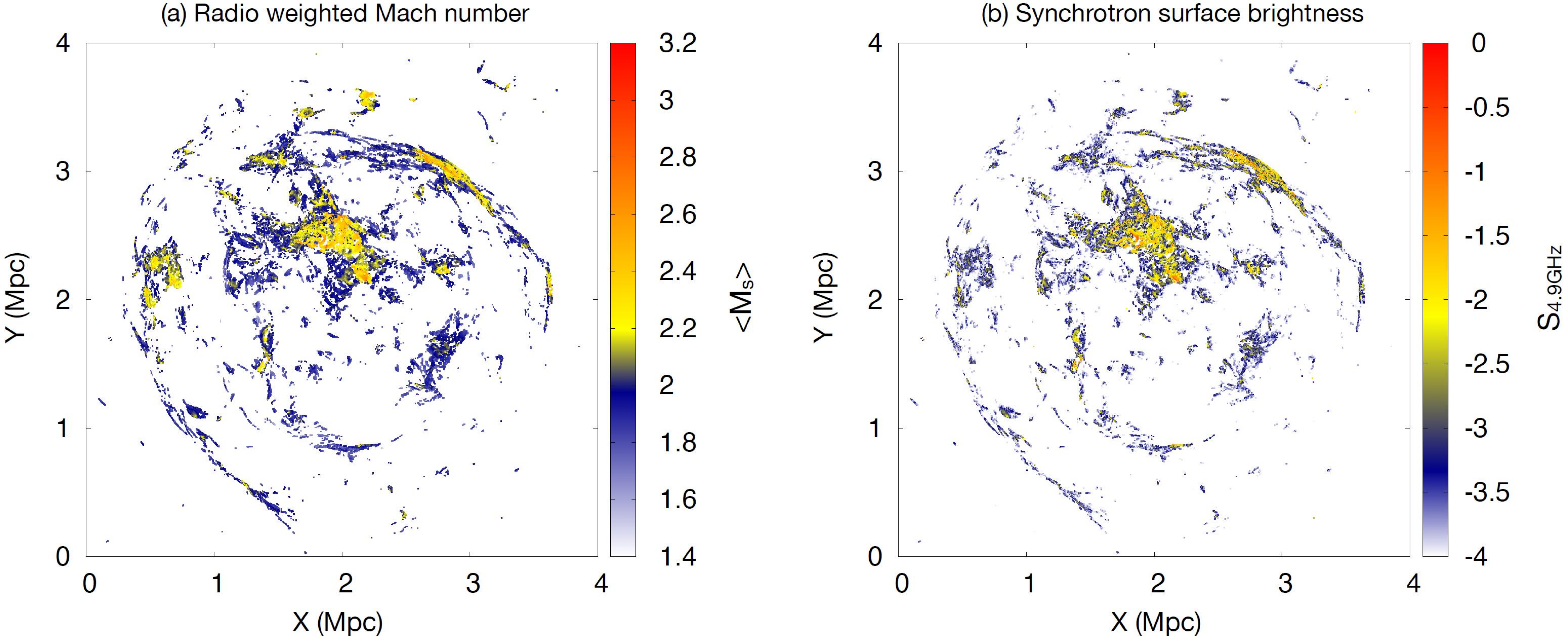}}
\vskip -0.2 cm
\caption{\label{fig:f6}
Maps projected along the line of sight ($z$-direction): (a) the shock Mach number weighted by synchrotron emissivity at 4.9 GHz, $\left<M_s\right>$,  and (b) the synchrotron surface brightness at 4.9 GHz, $S_{4.9}$, in the high-resolution simulation of $1024^3$ grid zones at $t=4.6$. The synchrotron emission was computed only at shock zones. See the text for the details of the calculations of $\left<M_s\right>$ and $S_{4.9}$.}
\end{figure*}

While the power spectrum in Figure \ref{fig:f4} tells us the scale of $B$ throughout the whole computational domain, the scale in clusters should differ at different radii from the center due to the density stratification. One way to examine the position-dependence of the scale of turbulence is through the so-called wavelet analysis \citep[e.g.,][]{farge1992,shi2018}. We calculated a wavelet power spectrum with the Mexican-hat wavelet. Then, the wavelet function is expressed as
\begin{equation}
\psi_{l,\bm{x'}}(\bm{x})\propto l^{-\frac{3}{2}}\left(3-\frac{|\bm{x}-\bm{x'}|^2}{l^2}\right)\exp\left(-\frac{|\bm{x}-\bm{x'}|^2}{2l^2}\right),
\end{equation}
where $l$ is the filtering scale. The wavelet coefficient of a fluid quantity, $f(\bm{x})$, is computed as
\begin{equation}
\hat{F}_l(\bm{x})=\int f(\bm{x'})\psi_{l,\bm{x'}}(\bm{x})d^3\bm{x'},
\end{equation}
and the wavelet power spectrum is given as $E_{\rm wavelet}(\bm{x})$ $\propto l^{-2}|\hat{F}_l(\bm{x})|^2$. 
Figure \ref{fig:f5} shows (a) the wavelet power spectrum, calculated as $l^{-2}(|\hat{V}_{xl}(\bm{x})|^2+|\hat{V}_{yl}(\bm{x})|^2+|\hat{V}_{zl}(\bm{x})|^2)$, for $\bm{v}$ and (b) the wavelet power spectrum calculated similarly for $\bm{B}$, as a function of $L_0/l$, at four different radii ($r_i= 0.2,~0.5,~1.0,$ and 1.5 Mpc), for the simulation same as in Figure \ref{fig:f4}(a). The power spectra averaged over a thin shell at the given radius are shown.

The wavelet power spectrum for $\bm{v}$ demonstrates that as we move from the core (red line) to the outskirt (yellow line), the relative fraction of the power of turbulent flow motions at large $l$, specifically, at $l\gtrsim L_0/10$, increases. The same trend was previously seen in cosmological structure-formation simulations \citep[e.g.,][]{shi2018}. Accordingly, the peak scale in the wavelet power spectrum for $\bm{B}$ also increases with the radius. While the peak is around $L_0/l\approx150$ at the core of $r=0.2$ Mpc, it is found around $L_0/l\approx100$ at the outskirt of $r=1-1.5$ Mpc. This indicates that the coherence length of $\bm{B}$ increases from the core to the outskirt, but the increment is expected to be less than a factor of two in our simulations.

\subsection{Turbulent Magnetic Fields in Radio Relics}\label{sec:3.3}

\begin{figure*}[t]
\vskip 0 cm
\hskip 0 cm
\centerline{\includegraphics[width=1.05\textwidth]{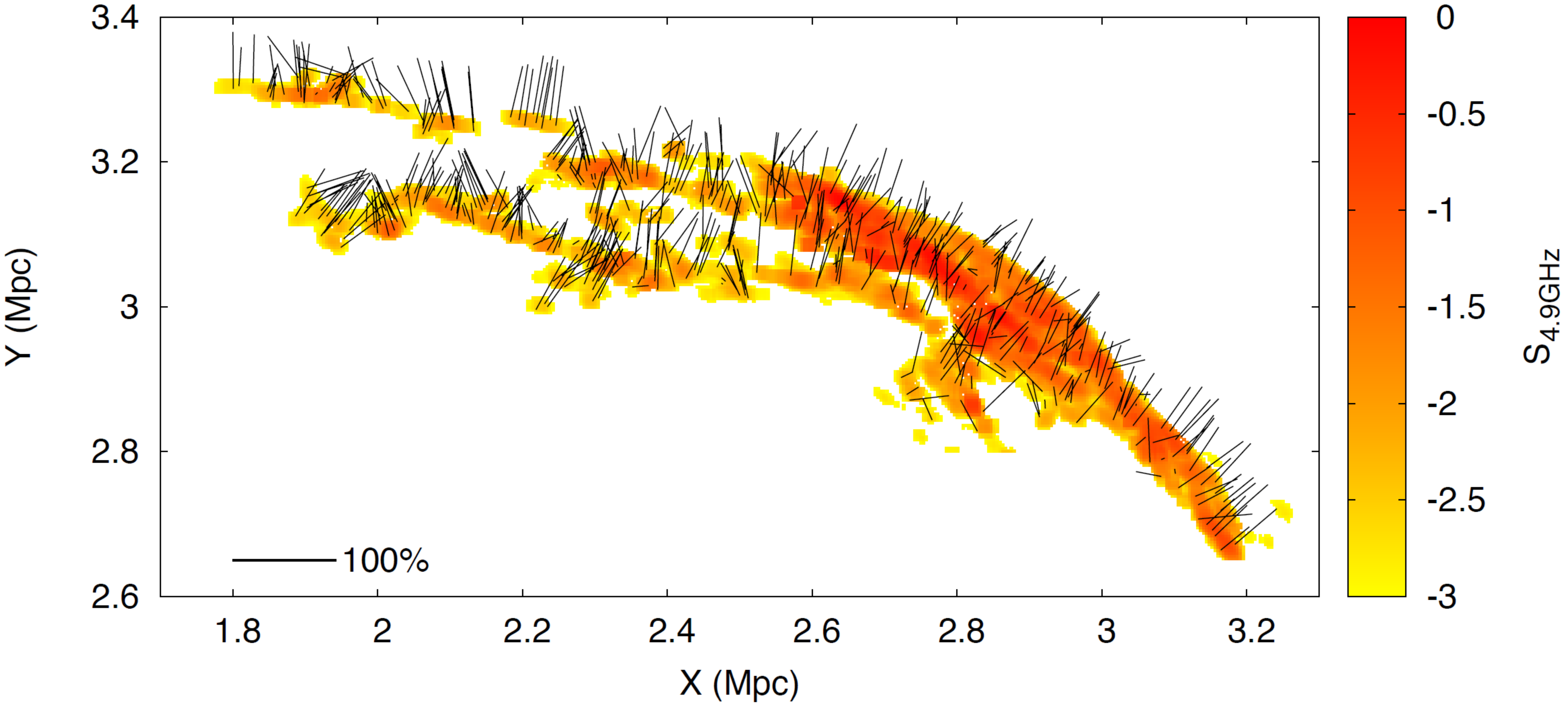}}
\vskip -0.2 cm
\caption{\label{fig:f7}
Zoom-in image of an upper-right part of Figure \ref{fig:f6}(b) for the synchrotron surface brightness, $S_{4.9}$, which mimics an observed giant radio relic. To match the resolution of radio observations, $S_{4.9}$ was smoothed over $\sim20\times20$ kpc$^2$. The black lines denote the polarization electric vectors, and their length scales with the polarization fraction.}
\end{figure*}

We point that the peak scale of $E_B(k)$ is $\sim200$ kpc and the most energy containing scale, that is, the peak scale of $kE_B(k)$, is $\sim65$ kpc in Figure \ref{fig:f4}(a). The peak of the wavelet power spectrum for $\bm{B}$ is at $l\approx40$ kpc in the cluster outskirt, as shown in Figure \ref{fig:f5}(b). All these are smaller than the size of giant radio relics by an order of magnitude or more (see the Introduction). It is not clear, however, how statistical quantities, like these peak scales, would be manifested in real observations. Hence, we here examine the structures of synchrotron polarization vectors in simulated radio relics, due to turbulence-generated magnetic fields, specifically including the effects of the shock compression of the magnetic fields.

To that end, we first identified shock zones (actually grid zones that are parts of shock surfaces) in the simulated cluster with the algorithm described in \citet{ryu2003} and \citet{pr2019}. The shock Mach number, $M_s$, was calculated using a formula from the jump condition for isothermal flows in Equation (9) of \citet{pr2019}. In fact, it is almost identical to that of hydrodynamic flows, $M_s\equiv\sqrt{\chi}$ (where $\chi$ is the density compression ratio across the shock), since the magnetic energy is substantially smaller than the kinetic energy in ICM shocks.\footnote{Most of identified shocks are fast shocks by the same reason.} 

We estimated the population of synchrotron-emitting cosmic-ray (CR) electrons as follow. CR electrons are assumed to be produced via diffusive shock acceleration (DSA) at ``quasi-perpendicular'' shocks with $\theta_{\rm Bn}\gtrsim45^{\circ}$, the obliquity angle between the upstream background magnetic field direction and the shock normal \citep[e.g.,][]{guo2014,kang2019}. The fraction of quasi-perpendicular shocks is $\sim70\%$ in our simulations, which is naturally expected with turbulence-generated ICM magnetic fields. The test-particle DSA model is assumed, because most of ICM shocks are weak with $M_s\lesssim3-4$ \citep[e.g.,][]{ryu2003,ha2018b}; then, the energy distribution of CR electrons can be modeled with a power-law form, $n_{\rm CR}=n_0\gamma^{-p}$ for $\gamma>\gamma_{\rm min}$, where $\gamma$ is the Lorentz factor and the DSA slope is $p=(2M_s^2+2)/(M_s^2-1)$ \citep[e.g.,][]{drury1983}. 

Although the detailed processes of CR electron acceleration at ICM shocks are yet to be understood \citep[e.g.,][]{guo2014,kang2019}, the acceleration efficiency $\eta_e$, the fraction of the shock energy transferred to CR electrons, is expected to increase with the shock Mach number. We adopted a rather simple scaling relation, $\eta_e\propto M_s$, for weak ICM shocks with $M_s\lesssim 4$. We also assumed that the electron acceleration is effective only in quasi-perpendicular shocks with $M_s\geq1.2$, and we choose $\gamma_{\rm min}=300$. The quantitative estimates of synchrotron emission by shock-accelerated CR electrons should depend on the details of the adopted model parameters. 
However, the main conclusion we will draw below should not be very sensitive to them, because we focus on the structures of polarization vectors of synchrotron emission, instead of the flux level.

We then calculated the synchrotron surface brightness. The synchrotron volume emissivity at frequency $\nu$ is given as $j_{\nu}\propto n_0 B_{\perp}^{(p+1)/2}\nu^{-(p-1)/2}$, where $B_{\perp}$ is the strength of the magnetic field component perpendicular to the line of sight (LoS). The Stokes parameters of the volume emissivity were calculated, using, for instance, $J_F(p)$ and $J_G(p)$ in Chapter 19 of \citet{shu1992}, in each shock zone with $B$ in the zone. The Stokes parameters of synchrotron surface brightness, $I_{\nu}$, $Q_{\nu}$, and $U_{\nu}$, were calculated by integrating the emissivity along LoSs, assuming that the diffuse ICM is optically thin to the synchrotron radiation.

Figure \ref{fig:f6} shows (a) the projected shock Mach number, $\left<M_s\right>$, weighted with the synchrotron emissivity at 4.9 GHz, along LOSs and (b) the synchrotron surface brightness at 4.9 GHz, $S_{4.9}$, in the $1024^3$ simulation at $t=4.6$. We note that the maps are produced with synchrotron emission only from shock zones; that is, the synchrotron radiation from the postshock zones behind the shock surface is not included. The grid size of this simulation is $\Delta x=3.9$ kpc, while the width of the synchrotron-emitting region behind the shock at $\nu=4.9$ GHz is $\Delta l_{4.9}\sim10$ kpc, if the postshock magnetic field strength is $\sim1$ $\mu$ G and the postshock flow speed is $\sim500$ km s$^{-1}$ \citep[e.g.,][]{kang2015}. Hence, the map of the surface brightness may not be the exact reproduction of observed radio relics. Yet, the figure shows that shocks formed in our simulations are weak with the projected Mach number of $\left<M_s\right>\lesssim3$. The synchrotron surface brightness reveals shell-like structures, as previously shown in full cosmological simulations \citep[e.g.,][]{skillman2013,wittor2017}.

The thin elongated part in the upper-right region of the shell-like structures in Figure \ref{fig:f6} looks similar to observed giant radio relics. It is zoomed in and shown in Figure \ref{fig:f7}; the surface brightness profile shown was smoothed over $4\times4$ grid zones, or $\sim20\times20$ kpc$^2$, to match the resolution of radio observations \citep[e.g.,][]{vanweeren2010,vanweeren2016}. 
The black lines represent the electric field vectors of polarized radiation; the fraction of linear polarization was calculated with $\sqrt{Q_{\nu}^2+U_{\nu}^2}/I_{\nu}$, and the angle $\chi$ was calculated with $\tan2\chi=U_{\nu}/Q_{\nu}$.

The average Mach number of the mock radio relic in Figure \ref{fig:f7} is $\sim2.3$, which is a bit smaller than, but close to, for instance, those inferred from the observed radio spectral index of the Sausage and Toothbrush relics \citep[e.g.,][]{vanweeren2016,hoang2017}. The projected Mach number, however, is not uniform over the whole structure, as can be seen in Figure \ref{fig:f6}(a). Rather it has a range of $\left<M_s\right>\sim1.6-2.5$. This is because the shock surfaces are not smooth, but composed of parts with different Mach numbers. This is in good agreement with the characteristics of ICM shocks formed in full cosmological structure-formation simulations \citep[e.g.,][]{hong2015,ha2018a}.

The polarization vectors of the mock radio relic look fairly organized, but not as well as those, for instance, in the Sausage and Toothbrush relics \citep[e.g.,][]{vanweeren2010,vanweeren2012}. The polarization fraction along the relic front is on average $\sim45\%$, which is smaller than that of the Sausage relic ($\sim50-60\%$) but a bit larger than that of Toothbrush ($\sim40\%$). However, the polarization angle between the polarization electric vectors and the shock normal is on average $\left<{\vartheta}_{\rm En}\right>\sim16^{\circ}$, about twice larger than that of the Sausage relic.\footnote{We estimated the angle for the Sausage relic by ourselves using the polarization vectors in Figure 3 of \citet{vanweeren2010}.} Although not shown here, we examined a number of mock radio relics from our simulations; the average polarization fraction is typically $\lesssim40\%$, and the average polarization angle is $\left<\vartheta_{\rm En}\right>\gtrsim20^{\circ}$. That is, our results indicate that with turbulence-generated magnetic fields in clusters, the highly organized structures of polarization vectors, such as those in the Sausage relic, would be difficult to be explained; the compression of the transverse components of turbulence-generated magnetic fields by weak ICM shocks may not be large enough to reproduce the observed levels of quasi-perpendicular obliquity angles of giant radio relics. This implies that for some observed radio relics, there may have been the pre-existing structures of magnetic fields, organized over the whole length scale of the radio relics.

We note that in the so-called reacceleration model for giant radio relics, ICM shocks are conjectured to sweep through the remnant bubbles of dead radio jets that contain fossil relativistic electrons with $\gamma \lesssim 300$ \citep[e.g.,][]{kang2012,kang2017}. This model was proposed mainly to explain the discrepancy between two kinds of shock Mach numbers inferred from radio and X-ray observations, and also to alleviate the problem of low acceleration efficiency in weak shocks and the low frequency of merging clusters with detected radio relics \citep{kang16}. However, if pre-existing magnetic fields are also required, their origin should be further investigated, but that is beyond the scope of this paper.

\section{Summary}\label{sec:summary}

Magnetic fields in galaxy clusters, at least in the outskirts, are conjectured to be originated by small-scale dynamo due to the turbulence induced during the hierarchical formation of the LSS of the universe. Yet, the characteristics of the dynamo has not been fully understood, because the turbulence in the ICM differs in some aspects from those in other astrophysical environments such as molecular clouds and star-forming regions. For instance, the turbulence is induced in the highly stratified ICM due to the gravity; also it is driven sporadically by mergers of sub-clusters. And most of all, the turbulence is subsonic with turbulence Mach number $M_{\rm turb}<1$. 

To estimate quantitative measures for turbulence and magnetic field amplification in the ICM, we performed a set of MHD simulations. Instead of running cosmological structure-formation simulations, we set up a model cluster with a radially stratified profile in a controlled periodic volume, and induced turbulence mimicking a series of sporadic merger events. A newly developed, high-order-accurate MHD code was employed. We then analyzed the characteristics of turbulence dynamo and the properties of resulting magnetic fields.

Our main findings can be recapitulated as follows.

1. Turbulence dynamo should be able to generate the cluster magnetic fields of $\sim\mu$G from the primordial seed fields of $\sim10^{-15}$ G or so within the age of the universe, if the MHD description of the ICM can be extended down to the scale of $\sim$ kpc or so.

2. With the compression of the transverse components of turbulence-generated magnetic fields by weak ICM shocks, some of the observed properties of synchrotron polarization in radio relics may be explained, but the highly organized structures of polarization vectors like those observed in the Sausage relic are difficult to be reproduced.

Finally, we note that the simulations described in this paper adopted an idealized setup with one model cluster in a periodic box, and hence did not include cosmological effects, such as the expansion of the universe and realistic mergers during the hierarchical formation of galaxy clusters. Hence, our findings should be eventually verified through full cosmological MHD simulations with high-resolutions of $\Delta x\sim 1$ kpc or so. Although such simulations are beyond the capacity of current supercomputing resources, they should be realizable in a near future.

\acknowledgments

We thank the anonymous referee for constructive comments that help us improve this paper. This work was supported by the National Research Foundation (NRF) of Korea through grants 2016R1A5A1013277, 2017R1A2A1A05071429, and 2017R1D1A1A09000567.

\end{document}